\documentclass[twocolumn,showpacs,preprintnumbers,superscriptaddress,amsmath,amssymb]{revtex4}
\usepackage{graphicx}% Include figure files
\usepackage{dcolumn}% Align table columns on decimal point
\usepackage{bm}% bold math
\begin{document}

%\markboth{Pisin Chen and Je-An Gu}
%{Dark Energy as a Manifestation of the Hierarchy}

\title{\bf Cosmological Constant as a Manifestation of the Hierarchy}

\author{Pisin Chen\footnote{The authors contribute to this work equally.}}
\email{chen@slac.stanford.edu} %
\affiliation{Kavli Institute for Particle Astrophysics and Cosmology, Stanford
Linear Accelerator Center, Stanford University, Stanford, CA
94305, U.S.A.} %
\affiliation{Department of Physics, Institute of Astrophysics and
Leung Center for Cosmology and Particle Astrophysics,
National Taiwan University, Taipei, Taiwan, R.O.C.} %

\author{Je-An Gu$^{*\,}$}
\email{jagu@phys.cts.nthu.edu.tw} %
\affiliation{National Center for Theoretical Sciences, Hsin-Chu, Taiwan, R.O.C.} %

\date{\today}

\begin{abstract}
There has been the suggestion that the cosmological constant as
implied by the dark energy is related to the well-known hierarchy
between the Planck scale, $M_{\rm Pl}$, and the Standard Model
scale, $M_{\rm SM}$. Here we further propose that the same
framework that addresses this hierarchy problem must also address
the smallness problem of the cosmological constant. Specifically,
we investigate the minimal supersymmetric (SUSY) extension of the
Randall-Sundrum model where SUSY-breaking is induced on the TeV
brane and transmitted into the bulk. We show that the Casimir
energy density of the system indeed conforms with the observed
dark energy scale.
\end{abstract}


% 11.10.Kk Field theories in dimensions other than four
% 11.25.Uv D branes
% 11.30.Pb Supersymmetry
%% 12.60.Jv Supersymmetric models
% 98.80.Es Observational cosmology
%          (including Hubble constant, distance scale, cosmological constant, early Universe, etc)

\maketitle

The accelerating expansion of the present universe was discovered in 1998
\cite{Perlmutter:1999np,Riess:1998cb} and further confirmed by type Ia
supernova (SN Ia) distance measurement \cite{SNLS,Riess:2006fw} and other
observations \cite{WMAP,Tegmark:2006}. This cosmic acceleration may be driven
by anti-gravity (repulsive gravity) generated by some energy source, generally
referred to as dark energy. By far positive cosmological constant (CC) is the
simplest realization of dark energy, which has become more favored by the
recent observations \cite{SNLS,Riess:2006fw,WMAP,Tegmark:2006}. If the dark
energy is indeed a cosmological constant which never changes in space and time,
then it must be a fundamental property of the spacetime. This would then
introduce a new energy scale, $M_{\rm CC}\simeq\rho_{\rm DE}^{1/4} \sim 10^{-3}
\,$eV, which is 15 orders of magnitude smaller than the Standard Model scale,
$M_{\rm SM} \sim {\rm TeV}$. Why is this energy gap so huge?

There has been another well-known hierarchy problem in physics,
i.e., the existence of a huge gap between the Standard Model scale
and the Planck scale of quantum gravity at $M_{\rm Pl} \sim
10^{19} \,$GeV by a factor $\sim 10^{16}$. The surprising
numerical coincidence between these two energy gaps prompts us to
the wonder: Are these two hierarchy problems related?

The idea that these two hierarchies are actually related is not
new. Arkani-Hamed et al.\ \cite{Arkani:2000} first invoked it to
address the cosmic coincidence problems. Various authors employed
it under different guises of ``cosmological constant seesaw
relation" \cite{McGuigan:2007}. Recently one of us introduced yet another variation of the theme \cite{Chen:2006}. If
one equates the two energy gaps as
\begin{equation}
M_{\rm CC}\simeq \frac{M_{\rm SM}}{M_{\rm pl}}M_{\rm
SM}=\Big(\frac{M_{\rm SM}}{M_{\rm Pl}}\Big)^2 M_{\rm Pl}
\equiv\alpha_G^2 M_{\rm Pl},
\end{equation}
then it suggests that the underlying mechanism which induces the CC
must be resulted from a double suppression by the same hierarchy
factor descended from the Planck scale to the SM scale.

We note that such a situation is not unique in physics. For
example, in atomic physics the hydrogen ground state energy is
suppressed from the electron rest mass, $m_e$, an energy scale
which enters naturally into the Schr$\ddot{o}$dinger equation, by
two powers of the fine structure constant, $\alpha=e^2/\hbar c$,
due to the presence of the coupling constant, $e$, in the Coulomb
potential. Analogous to that, the Planck-SM hierarchy ratio can be
viewed as a `gravity fine structure constant', $\alpha_G$. Various
physical energy scales in the system would then be associated with
the Planck mass through different powers of $\alpha_G$.

Guided by this philosophy, we construct a model for CC by
exploiting the Randall-Sundrum (RS1) \cite{RS} geometry that
addresses the SM-Planck hierarchy problem as our framework. To
accomplish our goal we find it necessary to extend the RS1 model to
incorporate minimal supersymmetry (SUSY). SUSY guarantees the
perfect cancellation of the vacuum energy between super-partners
from the outset. It therefore serves as a natural foundation in
solving the smallness problem of CC. Since SUSY must be broken in
our 4d world, we device its breaking via a Higgs field on the TeV
brane, which is then transmitted to the bulk through its coupling
to gravitino. Aside from this rather natural and minimal
extension, we follow the original RS1 scenario where the gravity
sector lives in the bulk while the standard model fields are
confined on the TeV brane. In the brane scenario where extra
dimensions are compactified the existence of the 4d Casimir energy
on the brane induced by the bulk field is inevitable. Such a
vacuum energy is a natural candidate for CC. Our task is to
demonstrate that the Casimir energy in our setup scales
generically as $\alpha_G^2 M_{\rm Pl}$.

Casimir effect has been considered as a possible origin for the
dark energy by many authors
\cite{Milton:2002hx,Gupta:2002,Bauer:2003,Aghababaie:2003wz,Burgess:2004kd,Chen:2004fn,Greene:2007}.
It is known that the conventional Casimir energy in the ordinary
3+1 dimensional spacetime cannot provide repulsive gravity
necessary for dark energy. On the other hand, Casimir energy on a
3-brane imbedded in a higher-dimensional world with suitable
boundary conditions can in principle give rise to a positive
cosmological constant. The general expression for the 4d Casimir
energy density on the 3-brane contributed from a bulk field is
given by
\begin{equation}
\rho^{(4)}_{\rm C} = \left[ \frac{1}{2} \int \frac{d^3p}{(2\pi)^3}
\sum_n \sqrt{p^2 + m_n^2} \right]_\textrm{ren} \, ,
\label{eq:Caimir energy density}
\end{equation}
where $p$ is the momentum in the ordinary 3-space, $m_n$ the mass
of the $n$th Kaluza-Klein (KK) mode, and the subscript ``ren''
stands for renormalization. For a massless field propagating in a
flat bulk, the KK mass scales as $m_n \sim n/a$, and the resulting
Casimir energy density on the brane scales as
\begin{equation}
\rho^{(4)}_{\rm C} \sim a^{-4}\, ,
\end{equation}
where $a$ is the extra dimension size. As summarized by Milton
\cite{Milton:2002hx}, for  $\rho^{(4)}_{\rm C}$ to
conform with dark energy the required extra dimension scale would have to be large.

Casimir energy in the RS1 geometry has been investigated by several
authors \cite{Naylor:2002,Setare:2003ds,Saharian:2003qs}. In the
supersymmetric brane-world, the contributions to the Casimir
energy on the brane from the bulk field superpartners
cancel each other perfectly. When SUSY is broken on the brane, its
modification of the KK mass spectrum becomes the primary source of
Casimir energy. Assuming that the gravitino KK mass spectrum is modified from
$m_n^2$ to $m_n^2 + \delta m_n^2$ while the graviton mass spectrum
remains unchanged, then the net 4d Casimir energy density on the
brane is
\begin{equation}
\Delta \rho^{(4)}_{\rm C} = \left. \rho^{(4)}_{\rm C}
\right|_{\delta m_n^2} - \left. \rho^{(4)}_{\rm C} \right|_{\delta
m_n^2 = 0 \; \forall \; n} \, ,
\end{equation}
where $\rho^{(4)}_{\rm C}|_{\delta m_n^2}$ denotes that in Eq.\
(\ref{eq:Caimir energy density}) with $m_n^2$ therein replaced by
$m_n^2 + \delta m_n^2$. Note that the term ``$-\rho^{(4)}_{\rm
C}|_{\delta m_n^2 = 0}$'' is exactly
the contribution from the SUSY partner. %
Generally speaking, when the SUSY-breaking induced KK mass-square
shift is much smaller than the energy gap, i.e., $\delta m_n^2 \ll
(m_n - m_{n-1})^2\equiv \Delta m_n^2$, and if both are insensitive to $n$,
%\begin{equation}
%\delta m_\textsc{susy}^2 \equiv \delta m_n^2\ll (m_n - m_{n-1})^2
%\equiv (\Delta m_\textsc{kk})^2 \, ,
%\end{equation}
%then it can be shown from Eq.\ (\ref{eq:Caimir energy density})
then it can be shown that the net 4d Casimir energy density on the
brane scales as
\begin{equation}
\Delta \rho^{(4)}_{\rm C} \sim \Delta m_{n}^2\delta m_{n}^2 \, . %
\end{equation}
We emphasize that this scaling for the brane Casimir energy under
SUSY-breaking is generic, and is essential for attaining the
desired CC scale. Relevant to our consideration, it has been shown
that in the RS1 geometry the graviton and gravitino KK mass
spectra on the TeV brane scales as $\Delta m_n\sim \alpha_GM_{\rm
Pl}$ \cite{GP1:2001a,GP2:2001b}, which is not surprising. What we
are obliged to demonstrate is that the graviton-gravitino KK
nonzero mode mass-square difference induced by our SUSY-breaking
mechanism is compelled to be as small as $\delta m_n^2 \sim
(\alpha_G^3 M_{\rm Pl})^2$. We will show that such an extremely
small value can emerge very naturally in our setup.

The Randall-Sundrum RS1 model invokes the following metric:
\begin{equation}
ds^2 = e^{-2\sigma} \eta_{\mu \nu} dx^{\mu} dx^{\nu} + a^2 dy^2\, ,
\label{eq:RS metric}
\end{equation}
where $\sigma=ka\vert y\vert$, $\mu,\nu=0,1,2,3$, $-\pi \leq y \leq \pi$, and
$a$ is the radius and $k$ the curvature of the orbifold
$\mathcal{S}^1/\mathcal{Z}_2$ in the compactified 5th dimension $y$. The
hidden, or Planck, brane locates at $y=0$ while the visible, or TeV, brane
locates at $y=\pi$. As is well-known, the Planck-SM hierarchy is bridged if
$ka\sim \mathcal{O}\left(10\right)$ so that the mass scale at $y=\pi$ is
suppressed by the warp factor $e^{-\pi ka}=\alpha_G$. It is customary to take
$k\sim M_{\rm Pl}$. So in the RS1 model the extra dimension size $a$ is only
about 10 times the Planck length.

Supersymmetry in a slice of AdS spacetime has been investigated by
various authors \cite{GP1:2001a,GP2:2001b,Altendorfer:2000}. The
complete supergravity action for our configuration would include
graviton, gravitino and graviphoton. But in our construction the
graviton KK masses would remain unchanged at the tree level in our
SUSY-breaking mechanism. Therefore it suffices our purpose that we
concentrate on the gravitino kinetic and mass terms only, which
are given by \cite{GP1:2001a}:
\begin{eqnarray}
S&=&S_5+S_0+S_{\pi} \, , \nonumber \\
S_5&=&\int d^4x\int dy \sqrt{-g}\Big[-\frac{1}{2}M_5^3\Big(\mathcal{R}
+i\bar{\Psi}^i_M\gamma^{MNP}D_N\Psi^i_P \nonumber \\
&&-i\frac{3}{2}\sigma'\bar{\Psi}^i_M\gamma^{MN}(\sigma_3)^{ij}\Psi^j_N\Big)-\Lambda\Big]
\, ,
\nonumber \\
S_{0,\pi}&=&\int d^4x\sqrt{-g_4}\Big[\mathcal{L}_{0,\pi}-\Lambda_{0,\pi}\Big]\,
.
\end{eqnarray}
Here $M,N,P=(\mu,5)$, $g={\rm det}(g_{MN})$, $\gamma^{MNP}$ is the
antisymmetric product of gamma matrices, and $\sigma'\equiv d\sigma/dy$. Note
that under our metric convention, $\sqrt{-g}=ae^{-4\sigma}$. $\mathcal{R}$ is
the 5d Ricci scalar, $\Psi^i_{M}$ $(i=1,2)$ the two symplectic Majorana
gravitinos, and $\Lambda$ is the bulk cosmological constant. $M_5$ is the 5d
Planck mass, which is related to the 4d Planck mass by $M_{\rm
Pl}^2=(1-\alpha_G)M_5^3/k.$

The RS1 metric solution of Eq.\ (\ref{eq:RS metric}) to the 5d
Einstein equations is valid provided that the bulk and boundary
cosmological constants are related by $\Lambda=-6M_5^3k^2\simeq
-6M_{\rm Pl}^2k^3$ and $\Lambda_{0}=-\Lambda_{\pi}=-\Lambda/k$.
%Supersymmetry automatically ensures the latter.
With $k\sim M_{\rm Pl}$, we have $\Lambda_{\pi}\simeq M_{\rm
Pl}^4$. This is associated with the ``old'' cosmological constant
problem \cite{Weinberg} where the field-theoretical argument would
result in the vacuum energy which is 123 orders of magnitude too
large. The resolution of this problem is beyond our ability here.
We merely assume that this problem would be resolved in the
future. (For a review of current attempts and possible approaches
to solving this problem, see \cite{Padmanabhan:2002ji}.)
%such as through the perfect cancellation between this surface
%tension on the brane and the pressure term from the bulk
%\cite{Chen:2000at}.
With the assumption that such an unphysical value for the vacuum
energy can be removed, our effort is to address the new
%, post dark energy
cosmological constant problem of its nonzero but extremely small
energy scale.

The gravitino supersymmetry transformation is given by
\begin{equation}
\delta\Psi^i_M=D_M\eta^i+\frac{\sigma'}{2}\gamma_M(\sigma_3)^{ij}\eta^j \, ,
\end{equation}
where the symplectic Majorana spinor $\eta^i (i=1,2)$ is the 5d supersymmetry
parameter. The equation of motion, i.e., the Rarita-Schwinger equation, for the
bulk gravitino in the AdS background is
\begin{equation}
\gamma^{NMP}D_N\Psi_P-\frac{3}{2}\sigma'\gamma^{MP}\Psi_P=0\, .
\end{equation}

The KK decomposition and the associated eigen-modes for bosons and fermions in
the RS1 geometry have been well studied in recent years
\cite{Goldberger:99,Flachi:2001,GP1:2001a,GP2:2001b}. Goldberger and Wise
\cite{Goldberger:99} first studied the behavior of bulk scalar field in the RS1
model. Flachi et al.\ \cite{Flachi:2001} investigated that for the bulk fermion
field. Gherghetta and Pomarol (GP1) \cite{GP1:2001a} extended the study to
different supermultiplets in the bulk. The bulk gravitino
field in the RS1 AdS geometry was studied in detail in a second paper by
Gherghetta and Pomarol (GP2) \cite{GP2:2001b}. Here we briefly summarize those
results relevant to our discussion. The 5d fields are decomposed as
\begin{equation}
{\rm \Psi}_{\mu,L,R}(x^{\mu},y)=\frac{1}{\sqrt{2\pi a}}\sum_n
\Psi^{(n)}_{\mu,L,R}(x^{\mu})f^{(n)}_{L,R}(y)\, ,
\end{equation}
where $\Psi_{\mu,L} (\Psi_{\mu,R})$ are defined even (odd) under the
$\mathcal{Z}_2$-parity. $\Psi_{5,L,R}$ and $\eta_{L,R}$ follow the similar KK
decomposition. GP2 solved the equation of motion and found the $y$-dependent KK
eigenfunction as
\begin{eqnarray}
f_L^{(n)} &=& \frac{1}{N_n}e^{3\sigma/2}\Big[J_{2}\Big(\frac{m_n}
{k}e^{\sigma}\Big) +bY_{2}\Big(\frac{m_n}{k}e^{\sigma}\Big)\Big],
\label{eq:fL} \\
f_R^{(n)} &=&
\frac{\sigma'}{kN_n}e^{3\sigma/2}\Big[J_{1}\Big(\frac{m_n}
{k}e^{\sigma}\Big) +bY_{1}\Big(\frac{m_n}{k}e^{\sigma }\Big)\Big],
\label{eq:fR}
\end{eqnarray}
where $m_n$ is the 4d mass for the $n$th mode and $b=-J_1(m_n/k)/Y_1(m_n/k)$
satisfies the boundary condition: $b(m_n)=b(\alpha_G^{-1}m_n)$. In the limit where $m_n\ll k$ and  $ka\gg 1$ the 4d KK masses for $n=1,2,...$, which are identical for both even and
odd modes, are found to be
\begin{equation}
m_n\simeq \alpha_G\Big(n+\frac{1}{4}\Big)\pi k \, .
\end{equation}
Note that the KK mass spectrum energy gap between
adjacent modes is independent of $n$, and $\Delta m_n\sim \pi\alpha_GM_{\rm Pl}\sim
{\rm TeV}$, as mentioned earlier.

Now we introduce the following action as a perturbation to break SUSY:
\begin{eqnarray}
S_{\Phi \Psi} &=&  \int d^4x\int \frac{dy}{a}\delta(y -\pi) \sqrt{-g} g_5
\Phi(x) \bar{\Psi}(x,y)
\Psi(x,y) \nonumber\\
                       &=& \sum_{n=0}^{\infty}\int \frac{dy}{a}\delta(y -\pi)
                        \sqrt{-g}\frac{1}{2\pi
                        a}[f^{(n)}]^2 \nonumber\\
                       &&\times\int d^4x g_5
                         \Phi(x)\bar{\Psi}^{(n)}(x)\Psi^{(n)}(x) \nonumber \\
                  &\equiv & \sum_{n=0}^{\infty}\int d^4x \delta m_n
                  \bar{\Psi}^{(n)}(x)\Psi^{(n)}(x)\, ,
\end{eqnarray}
where $\Phi$ is the fundamental Higgs field on the brane,
$f^{(n)}=f_L^{(n)}+f_R^{(n)}$ and $g_5$ the 5d Higgs-gravitino
Yukawa coupling. The gravitino mass-square shift for the $n$th KK
mode is thus
\begin{equation}
\delta m_n^2 = \left( g_5 \langle\Phi\rangle
                      \int \frac{dy}{a}\delta(y -\pi) \sqrt{-g}\frac{1}{2\pi
                      a}[f^{(n)}]^2\right)^2
                      \, ,
\end{equation}
where $\langle\Phi\rangle\sim M_{\rm Pl}$ is the vacuum
expectation value (vev) of the fundamental Higgs field [{\it cf.}
Eq.(17) in RS1]. As demonstrated in RS1, the physical mass scales
on the TeV brane are set, however, by the symmetry breaking scale,
$\alpha_GM_{\rm Pl}$, instead. The fundamental 5d coupling $g_5$
has the dimensionality of 1/mass and thus $g_5\sim 1/M_{\rm Pl}$.
Therefore $g_5\langle\Phi\rangle\sim 1$, and the suppression of
the SUSY-breaking gravitino mass shift is due solely to the
extreme smallness of the $y$-integral, which represents the
probability of finding the $n$th KK mode of gravitino on the TeV
brane.

We note that  for $n\geq 1$ the
argument of the Bessel functions in Eqs.\ (\ref{eq:fL}) and
(\ref{eq:fR}) is $\alpha_G^{-1}m_n/k \simeq (n+1/4)\pi \gg 1$. Accordingly
the values of the Bessel functions are either $\simeq
0$ or $\simeq \pm \sqrt{2/n\pi^2}$.
%%%%%%%%%%%%%%%%%%%%%%%%%%%%%%%%%%%%%%%%%%%%%%%%%%%%%%%%%%%%%%%%%%%%%%%
%\begin{eqnarray}
%\lim_{z\to \infty}J_{\nu}(z)&=& \sqrt{\frac{2}{\pi z}}\cos \left(
%z-\frac{\nu\pi}{2}-\frac{\pi}{4} \right)\nonumber\\
%&\simeq& \sqrt{\frac{2}{n\pi}}\cos[(n-1)\pi]=\pm
%\sqrt{\frac{2}{n\pi}}\, ,\nonumber\\
%%\end{eqnarray}
%%\begin(eqnarray}
% \lim_{z\to\infty}Y_{\nu}(z)&=&
%\sqrt{\frac{2}{\pi z}}\sin \left( z- \frac{\nu\pi}{2}-\frac{\pi}{4} \right) \nonumber\\
%&\simeq& \sqrt{\frac{2}{n\pi}} \sin[(n-1)\pi]=0\, .
%\end{eqnarray}
%%%%%%%%%%%%%%%%%%%%%%%%%%%%%%%%%%%%%%%%%%%%%%%%%%%%%%%%%%%%%%%%%%%%%%%
The normalization constant for the $n$th mode
can be determined from Eqs.\ (10)--(12) (via the normalization
condition for $\Psi_{\mu,L,R}$ \cite{GP1:2001a,GP2:2001b}), and it
can be shown that
\begin{equation}
N_n\simeq \frac{\alpha_G^{-1}}{\sqrt{2\pi
ka}}J_2(\alpha_G^{-1}m_n/k)\simeq
\frac{\alpha_G^{-1}}{\sqrt{n\pi^3 ka}}\, .
\end{equation}
It is interesting to note that in this limit the KK gravitino
wavefunction on the TeV brane is independent of $n$:
\begin{equation}
f^{(n)}(y=\pi)\simeq \sqrt{2 \pi k a} \, \alpha_G^{-1/2} \, .
\end{equation}
Collecting all the $\sigma$-dependences, we find that the $y$-integral of the
SUSY-breaking action scales as
\begin{equation}
\frac{1}{2\pi a}\int \frac{dy}{a}\delta(y -\pi) \sqrt{-g}[f^{(n)}]^2
\simeq k \alpha_G^3\, .
\end{equation}
Putting these together, we obtain the SUSY-breaking induced KK
gravitino mass-square shift for $n\geq 1$ modes:
\begin{equation}
\delta m_n^2 \sim \left( \alpha_G^3 M_{\rm Pl} \right)^2 .
\label{eq:delta mg}
\end{equation}
As for the $n=0$ mode, its wavefunction localizes on the Planck brane instead, with $f^{(0)}_L(y=\pi)\propto e^{-\sigma/2}=\alpha_G^{1/2}$ \cite{GP2:2001b}. This results in the mass shift $\delta m_0 \simeq \alpha_G^5M_{\rm Pl}$, which is totally negligible. 

Now we prove the generic Casimir energy scaling in
Eq.\ (5). 
%for our case where $m_n \simeq \alpha_G (n+1/4) \pi k$,
%$\Delta m_n \sim \alpha_G M_\textrm{Pl}$ and $\delta m_n^2 \sim
%(\alpha_G^3 M_\textrm{Pl})^2$. 
Employing the Jacobi $\theta$
function and the reflection formula,
\begin{equation}
\theta(z;x) = \sum_{n=-\infty}^{\infty} e^{-\pi n^2 x} e^{2\pi n
z} = \frac{1}{\sqrt{x}} e^{\pi z^2/x} \theta \left(
\frac{z}{ix};\frac{1}{x} \right)
\end{equation}
for the regularization, one can obtain
\begin{equation}
\rho^{(4)}_\textrm{C}(\mu^2) \sim - \frac{\Delta m_{n}^4}{2^5
\pi^{5/2}} \sum_{n=-\infty}^{\infty} \hspace{-0.6em}
\mbox{\raisebox{0.9ex}{$^{\prime}$}} \int_{0}^{\infty}
\frac{(-1)^n y^{3/2} dy}{\exp[\mu^2/y + 4 n^2 y]} \, ,
\end{equation}
where $\mu^2 \equiv \delta m_{n}^2 / \Delta m_{n}^2$ is
assumed to be insensitive to $n$, and the prime denotes the
summation without the $n=0$ term.

In the case where $\mu^2 \ll 1$,
%(from the Taylor expansion with respect to $\mu^2$ at $\mu^2 = 0$)
\begin{equation}
\rho^{(4)}_\textrm{C} (\mu^2) = \left. \rho^{(4)}_\textrm{C}
\right|_{\mu^2=0} + \left. \frac{d \rho^{(4)}_\textrm{C}}{d\mu^2}
\right|_{\mu^2=0} \mu^2 + \mathcal{O} (\mu^4) \, ,
\end{equation}
where
\begin{equation}
\left. \frac{d \rho^{(4)}_\textrm{C}}{d\mu^2} \right|_{\mu^2=0}
\sim \frac{\Delta m_{n}^4}{2^8 \pi^{5/2}}
\sum_{n=-\infty}^{\infty} \hspace{-0.6em}
\mbox{\raisebox{0.9ex}{$^{\prime}$}} \, \frac{(-1)^n}{n^{3}} \, .
\end{equation}
Thus,
\begin{equation}
\Delta \rho^{(4)}_\textrm{C} \simeq \left. \frac{d
\rho^{(4)}_\textrm{C}}{d\mu^2} \right|_{\mu^2=0} \mu^2 \sim \Delta
m_{n}^2 \delta m_{n}^2 \, .
\end{equation}
Consequently for our case
\begin{equation}
\Delta \rho^{(4)}_\textrm{C} \sim \alpha_G^8 M_\textrm{Pl}^4 \sim
\left[ \left( \frac{M_{\rm SM}}{M_{\rm Pl}} \right)^2 M_{\rm Pl}
\right]^4\sim M_{\rm CC}^4 \, .
\end{equation}

Recent observational evidence indicates that the dark energy may
actually be the cosmological constant. The surprising numerical
coincidence between the Planck-SM hierarchy and the SM-CC
hierarchy suggests a deeper connection between the two. In this
paper we explored the possibility of addressing these two
hierarchies within a single framework. Invoking the minimal SUSY
extension to RS1 model and SUSY-breaking on the TeV brane,
which is transmitted to the bulk through the Higgs-gravitino coupling,
we demonstrated that the 4d Casimir energy on the brane indeed
scales as $\alpha_G^2M_{\rm Pl}$, just right for the dark energy.
While the old cosmological constant problem is yet to be
addressed, it is remarkable that our model seems able to solve the
new CC problem rather naturally.

%\acknowledgements

\textbf{Acknowledgments}. This work is supported by the US
Department of Energy under Contract No.\ DE-AC03-76SF00515 and by
the Taiwan National Science Council under NSC 96-2119-M-007-001.


\begin{thebibliography}{00}
\bibitem{Perlmutter:1999np}
S.~Perlmutter {\it et al.}
%[Supernova Cosmology Project Collaboration],
%``Measurements of Omega and Lambda from 42 High-Redshift Supernovae,''
Astrophys.\ J.\  {\bf 517}, 565 (1999).
%%CITATION = ASTRO-PH 9812133;%%

\bibitem{Riess:1998cb}
A.~G.~Riess {\it et al.}
%[Supernova Search Team Collaboration],
%``Observational Evidence from Supernovae for an Accelerating Universe and a Cosmological Constant,''
Astron.\ J.\  {\bf 116}, 1009 (1998).
%%CITATION = ASTRO-PH 9805201;%%

\bibitem{SNLS}
  P.~Astier {\it et al.}  %[SNLS Collaboration],
  %``The Supernova Legacy Survey: Measurement of Omega_M, Omega_Lambda and w
  %from the First Year Data Set,''
  {\it A\&A} {\bf 447}, 31 (2006).
  %[arXiv:astro-ph/0510447].
  %%CITATION = AAEJA,447,31;%%

\bibitem{Riess:2006fw}
  A.~G.~Riess {\it et al.},
  %``New Hubble Space Telescope Discoveries of Type Ia Supernovae at $z > 1$:
  %Narrowing Constraints on the Early Behavior of Dark Energy,''
  arXiv:astro-ph/0611572 (2006).
  %%CITATION = ASTRO-PH/0611572;%%

\bibitem{WMAP}
  D.~N.~Spergel {\it et al.}  %[WMAP Collaboration],
  %``Wilkinson Microwave Anisotropy Probe (WMAP) three year results:
  %Implications for cosmology,''
  Astrophys.\ J.\ Suppl.\  {\bf 170}, 377 (2007).
  %[arXiv:astro-ph/0603449].
  %%CITATION = APJSA,170,377;%%

\bibitem{Tegmark:2006}
  M.~Tegmark {\it et al.},
  %``Cosmological Constraints from the SDSS Luminous Red Galaxies,''
  Phys.\ Rev.\  D {\bf 74}, 123507 (2006).
  %[arXiv:astro-ph/0608632].
  %%CITATION = PHRVA,D74,123507;%%

\bibitem{Arkani:2000}
N.~Arkani-Hamed, L.~J.~Hall, C.~F.~Kolda and H.~Murayama,
%``A New Perspective on Cosmic Coincidence Problems",
Phys. Rev. Lett. {\bf 85}, 4434 (2000).

\bibitem{McGuigan:2007}
See, for example, M.~McGuigan, arXiv:hep-th/0702182 (2007); 
V. V. Kiselev and . A. Timofeev, arXiv:0710.2204 [hep-th].

\bibitem{Chen:2006}
P.~Chen, Nucl.\ Phys.\ Proc.\ Suppl. {\bf 173} 137 (2007)
[arXiv:hep-ph/0611378].

%\bibitem{ADD}
%\bibitem{Arkani-Hamed:1998rs}
%N.~Arkani-Hamed, S.~Dimopoulos and G.~Dvali,
%``The hierarchy problem and new dimensions at a millimeter,''
%Phys.\ Lett.\ B {\bf 429}, 263 (1998).
%%CITATION = HEP-PH 9803315;%%

\bibitem{RS}
%\bibitem{Randall:1999ee}
L.~Randall and R.~Sundrum,
%``A large mass hierarchy from a small extra dimension,''
Phys.\ Rev.\ Lett.\ {\bf 83}, 3370 (1999).
%%CITATION = HEP-PH 9905221;%%

\bibitem{Milton:2002hx}
  K.~A.~Milton,
  %``Dark energy as evidence for extra dimensions,''
  Grav.\ Cosmol.\  {\bf 9}, 66 (2003).
  %[arXiv:hep-ph/0210170].
  %%CITATION = GRCOF,9,66;%%

\bibitem{Gupta:2002}
  A.~Gupta,
  %``Contribution of Kaluza-Klein modes to vacuum energy in models with large
  %extra dimensions and the cosmological constant,''
  arXiv:hep-th/0210069.
  %%CITATION = HEP-TH/0210069;%%

\bibitem{Bauer:2003}
  F.~Bauer, M.~Lindner and G.~Seidl,
  %``Casimir energy in deconstruction and the cosmological constant,''
  JHEP {\bf 0405}, 026 (2004)
  [arXiv:hep-th/0309200].
  %%CITATION = JHEPA,0405,026;%%

\bibitem{Aghababaie:2003wz}
  Y.~Aghababaie, C.~P.~Burgess, S.~L.~Parameswaran and F.~Quevedo,
  %``Towards a naturally small cosmological constant from branes in 6D
  %supergravity,''
  Nucl.\ Phys.\  B {\bf 680}, 389 (2004).
  %[arXiv:hep-th/0304256].
  %%CITATION = NUPHA,B680,389;%%

\bibitem{Burgess:2004kd}
  C.~P.~Burgess,
  %``Supersymmetric large extra dimensions and the cosmological constant: An
  %update,''
  Annals Phys.\  {\bf 313}, 283 (2004).
  %[arXiv:hep-th/0402200].
  %%CITATION = APNYA,313,283;%%

\bibitem{Chen:2004fn}
  P.~Chen and Je-An~Gu,
  %``Casimir effect in a supersymmetry-breaking brane-world as dark energy,''
  arXiv:astro-ph/0409238 (2004);
  P.~Chen and Je-An~Gu,
  %``A possible solution to the smallness problem of dark energy,''
  {\it eConf} {\bf C041213}, 1110 (2004).
  %%CITATION = ECONF,C041213,1110;%%

\bibitem{Greene:2007}
B. R. Greene and J. Levin,
%``Dark energy and stabilization of extra dimensions",
arXiv:0707.1062 [hep-th].

%\bibitem{rattazzi:2006} R.~Rattazzi, CERN-PH-TH/2006-029 [arXiv:hep-ph/0607055].

%\bibitem{Fabinger:2000} M. Fabinger and P. Horava, arXiv:hep-th/0002073 (2000).

%\bibitem{Randall:1999b} L. Randall and R. Sundrum, {\it Phys. Rev.
%Lett.} {\bf 83}, 4690 (1999).
%\bibitem{Gherghetta:2006} T. Gherghetta, {\it Les Houches Lecture
%on Warped Models and Holography}, arXiv:hep-ph/0601213 (2006).

\bibitem{Naylor:2002}
  W.~Naylor and M.~Sasaki,
  %``Casimir energy for de Sitter branes in bulk AdS(5),''
  Phys.\ Lett.\  B {\bf 542}, 289 (2002).
  %[arXiv:hep-th/0205277].
  %%CITATION = PHLTA,B542,289;%%

\bibitem{Setare:2003ds}
  M.~R.~Setare,
  %``Casimir energy densities for parallel plate on background of  conformally
  %flat brane-world geometries and cosmological constant  problem,''
  arXiv:hep-th/0308109.
  %%CITATION = HEP-TH/0308109;%%

\bibitem{Saharian:2003qs}
  A.~A.~Saharian,
  %``Wightman function and Casimir densities on AdS bulk with application to
  %the Randall-Sundrum braneworld,''
  Nucl.\ Phys.\  B {\bf 712}, 196 (2005).
  %[arXiv:hep-th/0312092].
  %%CITATION = NUPHA,B712,196;%%

%\bibitem{Birrell}
%N. D. Birrell and P.C.W. Davies, {\it Quantum Fields in Curved Space}, (Cambridge University Press, 1982).

\bibitem{GP1:2001a}
  T.~Gherghetta and A.~Pomarol,
  %``Bulk fields and supersymmetry in a slice of AdS,''
  Nucl.\ Phys.\  B {\bf 586}, 141 (2000)
  [arXiv:hep-ph/0003129].
  %%CITATION = NUPHA,B586,141;%%

\bibitem{GP2:2001b}
  T.~Gherghetta and A.~Pomarol,
  %``A warped supersymmetric standard model,''
  Nucl.\ Phys.\  B {\bf 602}, 3 (2001)
  [arXiv:hep-ph/0012378].
  %%CITATION = NUPHA,B602,3;%%

\bibitem{Altendorfer:2000}
  R.~Altendorfer, J.~Bagger and D.~Nemeschansky,
  %``Supersymmetric Randall-Sundrum scenario,''
  Phys.\ Rev.\  D {\bf 63}, 125025 (2001).
  %[arXiv:hep-th/0003117].
  %%CITATION = PHRVA,D63,125025;%%

\bibitem{Weinberg}
  S.~Weinberg,
  %``The cosmological constant problem,''
  Rev.\ Mod.\ Phys.\  {\bf 61}, 1 (1989).
  %%CITATION = RMPHA,61,1;%%

\bibitem{Padmanabhan:2002ji}
  T.~Padmanabhan,
  %``Cosmological constant: The weight of the vacuum,''
  Phys.\ Rept.\  {\bf 380}, 235 (2003).
  %[arXiv:hep-th/0212290].
  %%CITATION = PRPLC,380,235;%%

%\bibitem{Chen:2000at}
%  J.~W.~Chen, M.~A.~Luty and E.~Ponton,
%  %``A critical cosmological constant from millimeter extra dimensions,''
%  JHEP {\bf 0009}, 012 (2000)
%  [arXiv:hep-th/0003067].
%  %%CITATION = JHEPA,0009,012;%%

\bibitem{Goldberger:99}
  W.~D.~Goldberger and M.~B.~Wise,
  %``Bulk fields in the Randall-Sundrum compactification scenario,''
  Phys.\ Rev.\  D {\bf 60}, 107505 (1999)
  [arXiv:hep-ph/9907218].
  %%CITATION = PHRVA,D60,107505;%%

\bibitem{Flachi:2001}
  A.~Flachi, I.~G.~Moss and D.~J.~Toms,
  %``Quantized bulk fermions in the Randall-Sundrum brane model,''
  Phys.\ Rev.\  D {\bf 64}, 105029 (2001)
  [arXiv:hep-th/0106076].
  %%CITATION = PHRVA,D64,105029;%%


\end{thebibliography}
\end{document}